\documentstyle[aps]{revtex}
\begin{document}
\draft
\title{Systematics of the Quadrupole-Quadrupole Interaction and Convergence
Properties}
\author{M.S. Fayache$^{1}$, E. Moya de Guerra$^{2}$, P. Sarriguren$^{2}$, 
and L. Zamick$^{3}$}
\address{$^{1}$ D\'{e}partement de Physique, Facult\'{e} des Sciences \\
de Tunis, Tunis 1060, Tunisia\\
\noindent $^{2}$ Instituto de Estructura de la Materia, Consejo Superior de\\
Investigaciones Cient\'{\i }ficas, \\
Serrano 119, 28006 Madrid, Spain\\
$^{3}$ Department of Physics and Astronomy, Rutgers University, Piscataway,\\
New Jersey 08855}
\maketitle

\begin{abstract}
Our main concern in this work is to show how higher shell admixtures affect
the spectrum of a $Q\cdot Q$ interaction. We first review how, in the
valence space, the familiar SU(3) result for the energy spectrum can be
obtained using a coordinate space $Q\cdot Q$ interaction rather than the
Elliott one which is symmetric in ${\bf r}$ and ${\bf p}$. We then
reemphasize that the Elliott spectrum goes as $L(L+1)$ where $L$ is the
orbital angular momentum. While in many cases this is compatible with the
rotational formula which involves $I(I+1),$ where $I$ is the total angular
momentum, there are cases, e.g. odd-odd nuclei, where there is disagreement.
Finally, we consider higher shell admixtures and devise a scheme so as to
obtain results, with the $Q\cdot Q$ interaction, which converge as the model
spaces are increased. We consider not only ground state rotational bands but
also those that involve intruder states.
\end{abstract}

\section{Introduction}

There have been studies in the past of the quadrupole-quadrupole ($Q\cdot 
Q$) interaction in multishell spaces using the non-compact symplectic groups 
\cite{Goshen,Rosen}. Most of the focus of these studies has been on the
ground state rotational bands, the $\gamma $ vibration bands ($K=2$), the
giant quadrupole resonances and the effective charges. In this work we wish
to discuss the problems of extending the above studies to other topics such
as intruder states, odd-odd nuclei, etc. We feel that there has not been any
clear discussion of how higher shell admixtures affect the overall low lying
spectrum in a given nucleus, nor of how the low lying bands of odd-odd
nuclei are described with this interaction. Also we have recently shown how
to restore Elliott's SU(3) results \cite{Elliott} with a coordinate space
quadrupole-quadrupole interaction rather than one that is symmetric in
position and momentum. We will show that this is of more than academic
interest. We also discuss differences between Elliott's SU(3) approach \cite
{Elliott,EllHar,EllWill,Har} and the rotational model. The problem of
convergence with a $Q\cdot Q$ interaction is discussed in the last section.

The $Q\cdot Q$ interaction has a long history, the early part of which can
be found by reading excerpts from the text book by A.M. Lane \cite{lane},
and for recent applications the book by I. Talmi is recommended \cite{talmi}.

Following the notation of Golin and Zamick \cite{golin2} we define the
multipole-multipole interaction (including isospin) as

\begin{equation}
V\left( r\right) =-\chi \left[ \left( 2L+1\right) \left( 2T+1\right) \right]
^{1/2}\left[ Y^{L}\left( 1\right) Y^{L}\left( 2\right) \right] ^{0}f\left(
r_{1}\right) f\left( r_{2}\right) \left[ \delta _{T,1}\left[ \tau \left(
1\right) \tau \left( 2\right) \right] ^{0}+\delta _{T,0}\right]  \label{vr1}
\end{equation}
where $-\sqrt{3}\left[ \tau \left( 1\right) \tau \left( 2\right) \right]
^{0}=\tau \left( 1\right) \cdot \tau \left( 2\right) .$

The detailed expressions for the unnormalized antisymmetrized matrix
elements of this interaction are given in the above mentioned work of Golin
and Zamick \cite{golin2}. The expression is fairly complicated, involving
Racah coefficients. However the expression for the direct part of the
particle-hole matrix element (ph) for a ph state with total angular momentum 
$I$ and isospin $T^{\prime }$ is much simpler. The expression is
proportional to $\delta _{I,L}\delta _{T,T^{\prime }}$ and indeed there is
no other $I,T^{\prime }$ dependence. Thus, if one ignores the exchange part
of the interaction one has a simple schematic interaction. Thus, as Lane
describes \cite{lane}, we can use this interaction to describe collective
vibrational spectra.

The pairing plus quadrupole model was very popular and has been used by
Kisslinger and Sorensen \cite{kissl}, Baranger \cite{baranger} and Ikeda et
al. \cite{ikeda} around 1960, especially to describe the first excited 
$2^{+} $ state in vibrational nuclei. It can also be applied to giant
quadrupole resonances as emphasized by Bohr and Mottelson \cite{Bohr}. Here,
the schematic model of Brown and Bolsterli \cite{brobol} can be used. One
can also couple a particle to the giant quadrupole resonance and get an $E2$
effective charge. If one uses the selfconsistent strength for $\chi $ as
given by Bohr and Mottelson one finds that the isoscalar polarization charge 
$\delta e_{p}+\delta e_{n}$ is equal to the bare charge in an RPA
calculation.

In a different direction Elliott has shown \cite{Elliott} that a suitably
defined $Q\cdot Q$ interaction leads to rotational spectra in an open shell
nucleus with single-particle states defined by an isotropic
three-dimensional harmonic oscillator potential. His interaction is

\begin{equation}
V_{E}=-\frac{\chi }{2}\sum_{ij}\left( \frac{Q^{r}+Q^{p}}{2}\right) \cdot
\left( \frac{Q^{r}+Q^{p}}{2}\right)
\end{equation}
Note that it includes $i=j$ terms and that his quadrupole operator is
symmetric in position and momentum and this ensures that there will be no
mixing of major shells. This will be discussed in more detail later. Elliott
noted that the Casimir operator for the SU(3) group is

\begin{equation}
\tilde{C}_{2}=Q\cdot Q-3L\cdot L  \label{c2hat}
\end{equation}
From this he obtained the famous expression for the energy levels for a
rotational band for states with SU(3) quantum numbers $\lambda ,\mu $

\begin{equation}
\left\langle V_{E}\right\rangle =\bar{\chi}\left[ -4\left( \lambda ^{2}+\mu
^{2}+\lambda \mu +3\left( \lambda +\mu \right) \right) +3L\left( L+1\right)
\right]  \label{ener1}
\end{equation}
For fixed $\lambda ,\mu $ one gets the spectrum of a rotational band, but
going as $L\left( L+1\right) $ rather than $I\left( I+1\right) $.

We thus see that the $Q\cdot Q$ interaction has been used to make more
transparent how nuclear collectivity arises despite the fact that the shell
model with its implications of single particle motion seems to work. It
should be added that one can regard $Q\cdot Q$ as the long range part of a
more realistic interaction. This has been shown by many authors and we
recommend Talmi's book \cite{talmi} for a nice detailed derivation of this.

\section{Momentum dependent Q.Q interactions and the stability paradox}

It is of general interest to know how momentum dependent interactions affect
nuclear deformations. For example the Skyrme interaction contains the
momentum dependent terms $k^{\prime \;2}\delta \left( {\bf r}_{1}- 
{\bf r}_{2}\right) +\delta \left( {\bf r}_{1}-{\bf r}_{2}\right) k^{2}$ 
and ${\bf k}^{\prime }{\bf \cdot }\delta \left( {\bf r}_{1}-{\bf r}_{2} 
\right) {\bf k}$ \cite{Skyrme,Vautherin,Beiner}. 
The presence of those terms leads to an
effective mass $m^{*}<m$ in the nucleus. The rationale given to those
momentum dependent terms is that they simulate the finite range nature of
the nucleon-nucleon interaction. However, since the Skyrme interactions are
phenomenological they may be simulating more fundamental momentum dependence
of the nucleon-nucleon interaction.

The effect of the momentum dependent terms is known for the energy of the
isoscalar giant quadrupole resonance \cite{Golin,Bohr}

\begin{equation}
E^{*}=\sqrt{2}\hbar \omega \sqrt{\frac{m}{m^{*}}}\;.  \label{estar}
\end{equation}
The effect on the ground state deformations is not so well known.

Another example of momentum dependence is the Hamiltonian used by Elliott to
obtain SU(3) results \cite{Elliott}. This Hamiltonian is symmetric in
position and momentum coordinates

\begin{equation}
H=\sum_{i}\left( \frac{p^{2}}{2m}+\frac{1}{2}m\omega ^{2}r^{2}\right) -
\frac{\chi }{2}\sum_{ij}\left[ \frac{Q^{r}+Q^{p}}{2}\right] \cdot \left[ 
\frac{Q^{r}+Q^{p}}{2}\right] \;,  \label{hamil}
\end{equation}

\begin{equation}
Q_{\mu }^{r}=r^{2}Y_{2,\mu }\left( \Omega _{r}\right) ;\;Q_{\mu
}^{p}=b^{4}p^{2}Y_{2,\mu }\left( \Omega _{p}\right) ;\;\left( b^{2}=\hbar
/m\omega \right) \;.  \label{qdefs}
\end{equation}
We also define for convenience

\begin{equation}
Q^{E}=\frac{1}{2}\left( Q^{r}+Q^{p}\right) \;.  \label{qe}
\end{equation}

The momentum terms prevent $\Delta N=2$ mixing between major shells.
Although the original intention for this Hamiltonian is for valence nucleons
in a given major shell, let us use it here as an extreme example of a
momentum dependent Hamiltonian which is symmetric in ${\bf r}$ and ${\bf p}$
and perform a Hartree-Fock calculation with it.

In general, for any interaction in a deformed Hartree-Fock calculation the
condition for stability can be formulated as follows \cite{stabi}

\begin{equation}
\left\langle \sum_{i}\left( Q_{p}\right) \right\rangle =0\;,  \label{qp0}
\end{equation}
where $Q_{p}$ is the quadrupole moment operator in momentum space 

\begin{equation}
Q_{p}=b^{4}\sqrt{\frac{5}{16\pi }}\left(
2p_{z}^{2}-p_{x}^{2}-p_{y}^{2}\right) \;.  \label{qpdef}
\end{equation}

However, since the Hamiltonian is symmetric in ${\bf r}$ and ${\bf p}$, so
is the Wigner distribution function \cite{wigner}. This leads to the
conclusion $\left\langle \sum_{i}Q_{r}\right\rangle =0$, i.e., the
expectation value of the usual quadrupole moment also vanishes. But then
this goes against the common belief that the Elliott model supports
rotational motion.

A way out of this dilemma is to do, as we have done, go back to the ${\bf r}$
space interaction $Q^{r}\cdot Q^{r}$ and note that we can still get the
SU(3) results by including in the single-particle energies, not only the
contribution from the $i=j$ term in the sum 
\[
\frac{1}{2}\sum_{i,j}Q^{r}(i)\cdot Q^{r}(j) 
\]
but also the particle-core interaction \cite{Fayache,Moya}$.$ In these
references it was noted that $Q^{r}\cdot Q^{r}$ and $Q^{E}\cdot Q^{E}$ do
not give the same results in a valence space even if one uses harmonic
oscillator wave functions unless the particle-core interaction is included.
We can write

\begin{equation}
\frac{1}{2}\sum_{ij}Q\left( i\right) \cdot Q\left( j\right)
=\sum_{i<j}Q\left( i\right) \cdot Q\left( j\right) +\frac{1}{2}
\sum_{i}Q\left( i\right) \cdot Q\left( i\right) \;.  \label{qij}
\end{equation}
For the first term one does get the same result for $Q=Q^{E}$ and $Q=Q^{r}$
in a given $N$ shell. However this is not the case for the second $i=j$
term. Using $Q=Q^{r}$ one only gets $2/3$ of the value obtained with 
$Q=Q^{E} $. We have however shown \cite{Fayache,Moya} that the remaining 
$1/3$ comes from the (exchange) interaction of the valence nucleon with 
the core.

Just to clarify things, the diagonal part of $Q^{r}\cdot Q^{r}$ and the
particle-core interaction are the terms that give the single particle
splittings $\Delta _{N\left( \ell ,\ell ^{\prime }\right) }$ for different
$\ell $ values in a major shell. In order to get the SU(3) results of 
Elliott one must have such a single particle splitting as well as the 
two-body $Q\cdot Q$ interaction between the valence nucleons. The main 
difference then
between $Q^{E}\cdot Q^{E}$ and $Q^{r}\cdot Q^{r}$ is that with the former
all the single particle splitting comes from the $i=j$ part of $Q^{E}\cdot
Q^{E}$, whereas with $Q^{r}\cdot Q^{r}$ two thirds comes from the $i=j$ part
and one third from the particle-core interaction. In more detail the
expressions for the single particle splitting between $N\ell $ and $N\ell
^{\prime }$ states are \cite{Moya}

\[
\Delta _{N\left( \ell ,\ell ^{\prime }\right) }^{1}=2\bar{\chi}\left[ \ell
\left( \ell +1\right) -\ell ^{\prime }\left( \ell ^{\prime }+1\right)
\right] \;, 
\]

\[
\Delta _{N\left( \ell ,\ell ^{\prime }\right) }^{2}=\bar{\chi}\left[ \ell
\left( \ell +1\right) -\ell ^{\prime }\left( \ell ^{\prime }+1\right)
\right] \;, 
\]
where the $\Delta _{N\left( \ell ,\ell ^{\prime }\right) }^{1}$ comes from
the $i=j$ part of $Q^{r}\cdot Q^{r}$ and $\Delta _{N\left( \ell ,\ell
^{\prime }\right) }^{2}$ comes from the particle-core interaction.

\section{I(I+1) vs L(L+1) for rotational bands}

In this Section we stay in the valence space and make a comparison of
Elliott's SU(3) results \cite{Elliott,EllHar,EllWill,Har} with results of
the rotational model \cite{Bohr}. We feel that it has not been sufficiently
emphasized that one finds cases where these two models yield different
results for the behavior of rotational bands.

In the rotational model the formula for the energy of a state in a
rotational band with total angular momentum $I$ is given by \cite{Bohr}

\begin{equation}
E_{I}=E_{0}+\frac{\hbar ^{2}}{2{\cal I}}\left[ I\left( I+1\right) +\delta
_{K,1/2}\left( -1\right) ^{I+1/2}\left( I+1/2\right) a\right]  \label{erot}
\end{equation}
where $a$ is the decoupling parameter given by $a=-\left\langle K=1/2\left|
J_{+}\right| \overline{K=1/2}\right\rangle $ and where if $\left|
K\right\rangle =\sum_{j}C_{j,K}\phi _{j,K}$ then $\left| \bar{K}
\right\rangle =\sum_{j}C_{j,K}\left( -1\right) ^{j+K}\phi _{j,-K}$.

As noted by J.P. Davidson \cite{David} for an odd-odd nucleus with 
$K_{n}=\pm 1/2,\;K_{p}=\mp 1/2,\;\left( K=0\right) $, there is an additional
term $\left( -1\right) ^{I+1}a_{p}a_{n}\delta _{K,0}\left( \delta
_{K_{p},1/2}\delta _{K_{n},-1/2}+\delta _{K_{p},-1/2}\delta
_{K_{n},1/2}\right) $ (See also a recent review by Jain et al.\cite{Jain}).
For even-even nuclei, and for odd-even and even-odd nuclei with $K\neq 1/2$,
one gets the familiar $I\left( I+1\right) $ spectrum \cite{Bohr}.

It is generally thought that the Elliott SU(3) model also gives an $I\left(
I+1\right) $ spectrum. A careful reading of the papers however shows that
one really gets an $L\left( L+1\right) $ spectrum, where $L$ is the orbital
angular momentum \cite{Elliott,EllHar,EllWill,Har}. In this work we wish to
point out that to fully convey the similarities and differences of the
Elliott model and the rotational model, one should consider not only
even-even nuclei, but also even-odd (odd-even) and especially odd-odd
nuclei. The latter are usually not considered in the standard textbooks.

The point we make in this section is that the spectrum in the SU(3) model is
indeed $L\left( L+1\right) $ in all of the above instances. This leads to
several cases:

{\bf (a) Rotational bands where the spin }$S${\bf \ is equal to zero}. In
this case $I=L$ and one gets an $I\left( I+1\right) $ spectrum.

{\bf (b) }$K=1/2${\bf \ bands of even-odd (odd-even) nuclei. }Here again the
SU(3) spectrum is of $L\left( L+1\right) $ type. The Elliott formula is the
same as the rotational formula \cite{Goshen} including the decoupling term
evaluated in the asymptotic limit. Again, we have consistency between the
two approaches.

{\bf (c) Odd-odd nuclei. }Here again the SU(3) spectrum is $L\left(
L+1\right) $. There are some cases here where one gets a behavior which is
not consistent with the rotational formula.

We have performed shell model calculations with all possible configurations
in a given major shell using the interaction $\sum_{i<j}Q(i)\cdot Q(j)$
where, in order to get Elliott's SU(3) results we must, as mentioned in
Section II, also add single-particle splittings. Therefore, in the $1s-0d$
shell we have $\epsilon _{0d}-\epsilon _{1s}=18\bar{\chi}$ and in the $1p-0f$
shell we have $\epsilon _{0f}-\epsilon _{1p}=30\bar{\chi}$ , where 
$\bar{\chi}=\left( 5b^{4}/32\pi \right) \chi $ with $b$ the oscillator length
parameter $b^{2}=\hbar /m\omega $ \cite{Fayache,Moya}.

The same results can of course be obtained from the Elliott SU(3) formula
for the energies given in Eq.$\left( \ref{ener1}\right) $

One has the further rules \cite{Elliott,EllHar,EllWill,Har}:

Let $\bar{\lambda}$ be the maximum of $\lambda $ and $\mu $, and $\bar{\mu}$
the minimum. Then $K_{L}=\bar{\mu},\bar{\mu}-2,...,1$ or $0$ and

\begin{itemize}
\item  $L=K_{L},K_{L}+1,...,K_{L}+$ $\bar{\lambda}$ when $K_{L}\neq 0$;

\item  $L=\bar{\lambda},\bar{\lambda}-2,...,1$ or $0$ when $K_{L}=0.$
\end{itemize}

\subsection{A brief look at $K=1/2$ bands}

Let us be specific and discuss $^{19}$F and $^{43}$Sc. We consider in each
case three valence nucleons beyond a closed shell. In $^{19}$F the particles
are in the $1s-0d$ shell, whereas in $^{43}$Sc they are in the $1p-0f$
shell. The energy levels of the lowest bands are given in Table I for the
two cases. The results for the two nuclei are striking but different. In 
$^{19}$F, the lowest state is a $I=1/2^{+}$ singlet, and at higher energies
we get degenerate pairs 
$(3/2^{+},5/2^{+}),(7/2^{+},9/2^{+}),(11/2^{+},13/2^{+}).$ In $^{43}$Sc the
ground state is degenerate, and the degenerate pairs are 
$(1/2^{-},3/2^{-}),(5/2^{-},7/2^{-}),...,(17/2^{-},19/2^{-}).$

If we look at the rotational formula, we find that these results are
consistent with a decoupling parameter $a=+1$ for $^{19}$F and $a=-1$ for 
$^{43}$Sc. It is easy to show that these are precisely the $a-$values one
obtains with asymptotic Nilsson wave functions. In both cases the odd
particle will be in a $\Lambda =0,\Sigma =1/2$ state in the asymptotic
limit. From the definition of $\bar{K}$, the state $\left| \overline{\Lambda
=0\;\Sigma =1/2}\right\rangle $ can be shown to be equal to $-(-1)^{\pi
}\left| \Lambda =0\;\Sigma =-1/2\right\rangle $ , where $(-1)^{\pi }$ is
(+1) for an even-parity major shell and (-1) for an odd-parity one. Hence:

\begin{equation}
a=\left( -1\right) ^{\pi }\left\langle \Sigma =+1/2\left| J_{+}\right|
\Sigma =-1/2\right\rangle =\left( -1\right) ^{\pi }
\end{equation}

It has long ago been noted by Bohr and Mottelson \cite{Bohr} that $a=+1$
corresponds to weak coupling of the odd particle to $I=0^{+},2^{+},4^{+},...$
states, whereas $a=-1$ corresponds to weak coupling to $I=1,3,5$ states. In
the context of the SU(3) model, we would say that the $^{19}$F states have
an $L(L+1)$ spectrum with only even $L^{\prime }s$ allowed, and that 
$^{43}$Sc has an $L(L+1)$ spectrum with only odd $L^{\prime }s$ allowed.
It should be emphasized that the purpose of Table I is to compare the
results of the $Q\cdot Q$ interaction with the rotational model. To
compare with experiment,
more realistic interactions including in particular spin-orbit may be
required. In this respect, the results in Table I can be considered as the
asymptotic limit of large deformation.

At any rate, we have shown that the $Q\cdot Q$ interaction gives the same
results for these two $K=1/2$ bands as does the rotational formula with
asymptotic Nilsson wave functions.

Although it is not our intention in this work to fit experiment, rather we
wish to work out the consequences of our model, we cannot resist discussing
an interesting result for $^{19}$F. There is a beautiful example of weak
coupling in $^{19}$F corresponding to $a=+1$, but it is not the positive
parity band discussed above but rather a negative parity band corresponding
to a 4-particle 1-hole configuration. This has been discussed by Nazarewicz
et al. \cite{Naza}. In the asymptotic limit the proton hole would be in the
Nilsson orbit $\left[ 101\right] \frac{1}{2}$ which would yield $a=0$.
However, the dominance of the $p_{1/2}$ due to spin-orbit interaction causes 
$a$ to be very close to one. The spectrum looks like the weak coupling of
a $p_{1/2}$ hole to the states of the ground state rotational band of 
$^{20}$Ne.

\subsection{Odd-odd nuclei, e.g. $^{22}$Na}

In Table II we show a fairly detailed list of energy levels for the odd-odd
nucleus $^{22}$Na obtained with the $Q\cdot Q$ interaction. We show $T=0$
and $T=1$ states in separate columns. We use the same parameters as in 
$^{19} $F just to bring out some similarities. If one is interested in a 
best fit, one should of course have an $A$ dependence in $\chi $.

An striking feature in this table is that many states with diverse $I^{\pi
}T $ assignments are degenerate. This is due to the prevailing $SU(4)\otimes
SU(3)$ symmetry. We shall come back to this point at the end of this section
but first we compare with the rotational model.

We have underlined $T=0$ and $T=1$ rotational bands, and will now discuss
them in more detail. Note that the ground state consists of two degenerate
states, one with $I=1^{+}\;T=0$ and the other with $I=0^{+}\;T=1$. Both
states have $L=0$ and the simple spin-independent interaction gives the same
energy for $S=0$ and $S=1$. Let us first look at the underlined $T=1$
states. The ground state is $I=0^{+}$, the $2^{+}$state is at 1.588, the 
$4^{+}$ is at 5.293, etc. If we follow the rotational sequence 
$I=0^{+},2^{+},4^{+},...$ we see a simple rotational behavior:

\[
E(I)-E\left( 0_{1}^{+}\right) =AI\left( I+1\right) ;\hspace{2cm}
\text{ with } A=\frac{\hbar ^{2}}{2{\cal I}}=\frac{E\left( 2^{+}\right) 
-E\left( 0_{1}^{+}\right) }{6} 
\]
There is nothing new here.

The excitation energies of the underlined 
$T=0,\;I=1^{+},2^{+},3^{+},...,10^{+}$ states follow the sequence

\[
E^{*}(I)\equiv E(I)-E(1_{1}^{+})=AI(I+1);\hspace{2cm}\text{ with }A=
\frac{E\left( 2^{+}\right) -E\left( 1_{1}^{+}\right) }{6} 
\]
However the rotational model formula (Eq. (\ref{erot})) would give a
different energy level spacing

\[
E^{*}(I)\equiv E(I)-E(1_{1}^{+})=A^{\prime }\left[ I(I+1)-2\right] ;
\hspace{2cm}\text{ with }A^{\prime }=\frac{E\left( 2^{+}\right) -E\left(
1_{1}^{+}\right) }{4}, 
\]
which is not followed by the above mentioned energy levels in the table.
Thus, for the case of $T=0$ states in odd-odd nuclei we get a clear
difference between the rotational formula and SU(3).

We gain further insight by examining the degeneracies associated with the 
$T=0$ underlined states in Table II, i.e., those with energy $AI(I+1)$. The
even $I$ states up to $I=8$ are doubly degenerate whereas the $I=10$ and the
odd-$I$ states are singlets. This suggests that there are two bands for
which the states with the same $I$ values are degenerate. One band is a $K=2$
band with all values of $I$ from 2 to 10, and there is nothing anomalous
about it. The other band consists of states of angular momentum 1,2,4,6 and
8. For the latter band, the orbital angular momentum of the states are
0,2,4,6 and 8 respectively, and they all have $S=1.$ Their energies can be
fit to the formula $E^{*}(I)=AL(L+1)$ rather than $AI(I+1)$, such that only
even $L$ contribute.

Let us now discuss on the basis of SU(3) the degeneracy observed in Table
II, limiting the discussion to $T=0$ and $T=1$ states (similar arguments
would follow for higher isospin states). For $^{22}$Na we have 3 protons and
3 neutrons in the $N=2$ shell. The ground state will correspond to maximal
spatial symmetry in coordinate space, i.e., to the $\left( \lambda ,\mu
\right) =\left( 8,2\right) $ representation of the $\left[ f\right] =\left[
4,2\right] $ partition. The possible $K_{L},L$ values are

\begin{itemize}
\item  $K_{L}=0:\;L=0,2,4,6,8$

\item  $K_{L}=2:\;L=2,3,4,5,...,10$
\end{itemize}

Since the energy does not depend on $K_{L}$ (see Eq.(\ref{ener1})), states
with equal $L-$values and different $K_{L}$ are degenerate. In addition,
antisymmetry in spin-isospin space demands that $S=0$ when $T=1$, and $S=1$
when $T=0$. Therefore, $T=1$ states with $I=L,$ and $T=0$ states with 
$I=L,L+1,L-1$ will be degenerate. Thus for the ground state $\left[
4,2\right] \left( 8,2\right) $ representation we find with the same
excitation energy $E^{*}\left[ L\right] =3\bar{\chi}L\left( L+1\right) ,$
the following states:

\begin{itemize}
\item  For $L=2,4,6,8$

\begin{itemize}
\item  two states, each with $T=0$ and $I=L,L\pm 1,$ and

\item  two states, each with $T=1$ and $I=L,$
\end{itemize}

\item  For $L=3,5,7,9,10$

\begin{itemize}
\item  one state with $T=0$ and $I=L,L\pm 1,$ and

\item  one state with $T=1$ and $I=L.$
\end{itemize}
\end{itemize}

This explains why the state with excitation energy $1.588$ MeV (i.e., $L=2$)
appears twice for $T=0$ and $I=1,2,3$ and twice for $T=1$ and $I=2$. The
state with excitation energy $3.176$ MeV (i.e., $L=3$) appears once with 
$T=0,I=2,3,4$ and once with $T=1,I=3$. The next state of this representation
is at $5.293$ MeV (i.e., $L=4$), and according to the above discussion
should appear twice for $T=0$ and $I=3,4,5$ and twice for $T=1$ and $I=4$.

Surprisingly though there are more degeneracies at $5.293$ MeV, namely one
$T=0$ state with $I=3$ and three $T=1$ states with $I=2,3,4$. This is due to
an additional degeneracy of $\left[ 4,2\right] \left( 8,2\right) L=4$ with
the $\left[ 4,1,1\right] \left( 9,0\right) L=3$ state that corresponds to
total symmetry in spin-isospin space (i.e., $T=0,S=0$ or $T=1,S=1$). The
allowed $K_{L},L$ values in this representation are $K_{L}=0,L=1,3,5,7,9,$
and the excitation energy for a given $L-$value, $E^{*}\left[ (90)L\right]
-E_{GS}^{*}\left[ 0\right] =\bar{\chi}\left[ 24+3L\left( L+1\right) \right]
, $ is shared by states with $T=0,I=L$ and with $T=1,I=L,L\pm 1.$ Thus the
states that belong to this representation for $L=1$ are the states at $2.647$
MeV with $T=0\;I=1$ and $T=1\;I=0,1,2,$ while for $L=3$ we get the states at 
$5.294$ MeV previously discussed with $T=0\;I=3$ and with $T=1\;I=4,$ plus
the states $T=1\;I=2,3.$ The next states with $L=5\;\left(
E^{*}=10.055\right) $ appear for $T=0\;I=5,\;T=1\;I=4,5,6,$ and are
degenerate with lower angular momentum states corresponding to $K_{L}=1,L=2$
in the $\left( 6,3\right) $ representation (The $L=1$ states in this last
representation have 9 MeV excitation energy). By the same token one can
explain the remaining energy levels in the table. To the best of our
knowledge there has been no discussion previously of these additional
degeneracies corresponding to different values in different representations.

Experimentally the lowest band in $^{22}$Na has $K=3$. In the Nilsson scheme
when we fill the lowest $N=2\;K=1/2$ level we reach $^{20}$Ne. To form $^{22}
$Na we put the odd neutron and odd proton into the $K=3/2$ level and the
Gallagher Moszkowski rule favors $K=3$ over $K=0$ for the lowest band \cite
{Galla}. However, with pure $Q\cdot Q$ as seen in Table II, the lowest 
$I=3^{+}$ state comes out at 1.588 MeV of excitation. Clearly, the
limitations of the SU(3) formula for energy levels are more apparent when
one examines the spectra of odd-even and odd-odd nuclei rather than limiting
oneself to even-even nuclei. The main deficiency is the absence of a
spin-orbit interaction in the SU(3) model. For even-even nuclei the lowest
lying levels are dominantly $S=0$ states but this is not the case for $T=0$
states of odd-odd nuclei.

\section{Convergence with a Q.Q interaction}

We now come to the last topic of this work -- how to achieve convergence
with a $Q\cdot Q$ interaction. When we remove the momentum dependent terms
from $Q\cdot Q$ and go back to the ${\bf r}$ space $Q^{r}\cdot Q^{r}$, we
can mix major shells. Therefore, this coordinate space interaction combines
the advantages of the algebraic Elliott's model, with the possibility of
studying intruder states and admixtures of higher shells in the ground state.

There is a general belief that if we go to larger and larger spaces the
results will diverge. This is because the $Q^{r}\cdot Q^{r}$ interaction
increases its strength as the distance between two nucleons increases.
However, we wish to show in this work that we can get around this problem.
Just as people do all the time in G-matrix calculations, we can modify the
interaction as we change the model space. If we allow up to $n\hbar \omega $
excitations in the model space we can make the strength of the $Q^{r}\cdot
Q^{r}$ depend on $n$

\[
V=-\frac{\chi _{n}}{2}Q^{r}\cdot Q^{r}\;. 
\]

A simple scheme for getting convergence is to demand that the energy of
the $I=2_{1}^{+}$ state in an open shell nucleus comes out correctly (i.e.,
agrees with experiment) for each $n$. We have applied this scheme to the 
$I=0^{+},1^{+}$ and $2^{+}$ (all $T=0$) states in $^{8}$Be. These are shown
in Tables III, IV, and V, respectively. In each table the results are given
in a $0\hbar \omega $ space $\left( n=0\right) $, $(0+2)\hbar \omega $ space 
$\left( n=2\right) $, and ($0+2+4)\hbar \omega $ space $\left( n=4\right) $.
In order to get the $I=2_{1}^{+}$ state to come out correctly we had to
steadily decrease $\chi $ with increasing $n$. The values for $n=0,2,$ and 4
are respectively $0.5216,\;0.4119,$ and $0.3369\;$MeV\ fm$^{-4}.$

The calculations have been performed using the OXBASH program \cite{bro}. In
this program the effects of spurious states are removed using the
Gloeckner-Lawson method \cite{gloec}. In this method the spurious states are
pushed up to a very high energy. This is done by diagonalizing the modified
Hamiltonian for a system of A nucleons

\begin{equation}
H^{\prime }=H_{SM}+\lambda _{CM}\left[ \frac{{\bf P}^{2}}{2Am}+
\frac{1}{2}mA\omega ^{2}{\bf R}^{2}-\frac{3}{2}\hbar \omega \right]  
\label{hprime}
\end{equation}
where $H_{SM}$ is the usual shell model Hamiltonian and ${\bf P}$ and 
${\bf R}$ are the position vectors for the center of mass. By making 
$\lambda _{CM}$
very large, the spurious states corresponding to center of mass motion will
be pushed up to a very high energy. They will be well separated from the
lower lying physical states.

Let us first discuss the calculations up to $2\hbar \omega $, the results of
which are given in the tables III, IV and V. These excitations are of two
types, we can either excite one particle through 2 major shells $\left(
1p-1h\right) $ or we can excite two nucleons through one major shell $\left(
2p-2h\right) $. The $\left( 1p-1h\right) $ excitations can be regarded as
giant resonance states built upon the valence states as have indeed been
studied in the context of symplectic symmetry \cite{Rosen}. The giant
resonance states are generally at high excitation energies but their
admixtures into the valence states can lead to important effects. For
example, the isoscalar $E2$ effective charge needed to fit experiments in
valence space calculations is about a factor two larger than the bare charge
and this can be understood in terms of small admixtures of the $2\hbar
\omega $ $\left( 1p-1h\right) $states into the basis valence states.

We next consider the $2p-2h$ states. In some nuclei some states of this type
come much lower in energy than the $2\hbar \omega $ estimate. They may
intertwine with low lying valence states. In such cases these $2p-2h$ states
or more generally, $np-nh$ states, are called {\it intruder states}. In
previous calculations \cite{Faya98} we have shown that there are such low
lying intruders in $^{10}$Be, $^{12}$C and $^{16}$O, but actually not in 
$^{8}$Be.

The reason that in some nuclei the intruders come low in energy and in
others they do not, can be understood in the context of the Nilsson model.
The optimum way to have a low lying intruder is to lift two nucleons from a
Nilsson orbit whose energy increases with deformation (upgoing level) and to
put them into a Nilsson orbit whose energy decreases with deformation
(downgoing level). For $^{10}$Be, $^{12}$C and $^{16}$O we can remove
nucleons from upgoing levels in the $0p-$shell, but in the case of $^{8}$Be
the valence levels are depleted and we would have to remove nucleons from a
downgoing level.

Before discussing the energy levels we will make some comments about the
occupancies. It had been previously noted by Fayache et al. \cite{Faya97}
that if one allows only $2\hbar \omega $ admixtures with the $Q\cdot Q$
interaction then the shell model matrix reduces into two parts. First, we
have states which are admixtures of valence states (no particles excited to
higher shells) and states in which one nucleon is excited through two major
shells, and second we have states in which two particles are excited through
one major shell. There is no mixing in between the $\left(
0p-0h+1p-1h\right) $ states and the $2p-2h$ states. This is because of the
'parity' selection rule that $Q\cdot Q$ cannot excite two nucleons from an
even (odd) parity major shell to an odd (even) major shell.

If we look at Tables III, IV, and V under the columns $\left( 0+2\right)
\hbar \omega $ we see several states which have 100\% $2\hbar \omega $
configurations. These are precisely the $2p-2h$ states which are uncoupled
from the rest. When we allow $4\hbar \omega $ admixtures the $2p-2h$ states
admix with states in which one nucleon is excited through one major shell
and the other through three major shells. We therefore no longer get 100\%
$2\hbar \omega $ for these states, but they are still uncoupled from states
which have mainly valence configurations.

In more detail, for $I=0^{+}$ in Table III, in the $\left( 0+2\right) \hbar
\omega $ column the ground state has 21.77\% $2\hbar \omega $ admixtures.
The next four states have smaller $2\hbar \omega $ admixtures, but the state
at 30.951 MeV has an 88.89\% admixture. This is clearly a state which is
dominantly a $1p-1h$ state in which a nucleon has been excited through 2
major shells. This state can be easily distinguished from the $2p-2h$ states
at 32.214 MeV and 34.409 MeV. The latter have 100\% $2\hbar \omega $
configurations. As mentioned before there is no mixing, at the $2\hbar
\omega $ level between the $\left( 0p-0h+1p-1h\right) $ states and the 
$2p-2h $ states via a $Q\cdot Q$ interaction.

States with 100\% occupancy are also seen in Tables IV and V corresponding
to $I=1^{+}$ and $2^{+}$. These all lie in the energy range 30-35 MeV, and
are clearly $2p-2h$ states. As mentioned before there are no low lying
intruders in $^{8}$Be. These $2p-2h$ states are close in energy with the
(dominantly) $1p-1h$ states for $^{8}$Be.

When we go to the $\left( 0+2+4\right) \hbar \omega $ column the results are
not so striking -- we do not see 100\% of any configuration. But we can
still distinguish the $\left( 0p-0h+1p-1h\right) $ states from the $\left(
2p-2h+2p-2h\otimes 1p-1h\right) $ states. For the latter the sum of the 
$2\hbar \omega +4\hbar \omega $ occupancy should be 100\%. For the $I=0^{+}$
states in Table III this is not the case for the state at 26.225 MeV
(71.5\%) but it is the case for states at 32.085 MeV and 34.200 MeV. Similar
states can be found for $I=1^{+}$ and $2^{+}$, i.e. states where the $2\hbar
\omega +4\hbar \omega $ occupancy is 100\%. This confirms that with $Q\cdot
Q $ one gets a separation of different classes of states.

We now discuss the energy levels as a function of $n$, the number of $\hbar
\omega $ excitations. While there are rather large deviations in going
from $n=0$ to $n=2$, there is excellent convergence for many states in
comparing $n=2$ and $n=4$. The percentage deviation for the $I=0^{+}$ case
for the first 4 states (up to about 20 MeV) are respectively 
$0.77,\;2.22,\;2.75,$
and $3.93\%$. For the next state, which is at 26.225 MeV in the 
$(0+2+4)\hbar \omega $ calculation, the percentage deviation is large 
$18.02\% $. But if we examine this state we see that it is mainly a $2\hbar
\omega $ excitation state, not present in the $0\hbar \omega $ calculation.
It corresponds to a $1p-1h$ excitation through 2 major shells, as opposed to
many other 2$\hbar \omega $ states which correspond to two nucleons being
excited through one major shell.

We see that the convergence between $\left( 0+2\right) \hbar \omega $ and 
$\left( 0+2+4\right) \hbar \omega $ holds to surprisingly high energies. For
example for $I=1^{+}$ at the $\left( 0+2\right) \hbar \omega $ energy level
there are five consecutive $2p-2h$ states between 30.722 MeV and 34.503 MeV.
The \% deviations in the energies relative to the $\left( 0+2+4\right) \hbar
\omega $ calculations are respectively 10.13, 9.19, 4.84, 2.01 and 1.87\%.

\section{Acknowledgments}

This work was supported by the U.S. Department of Energy Grant No.
DE-FG02-95ER-40940 and by DGICYT (Spain) under Contract No. PB95/0123. We
thank Ben Bayman and John Millener for useful comments and insights, and
Durga Devi for her help.

\nopagebreak

\newpage

\widetext

\begin{table}[tbp]
{\bf Table I. }Energy levels (MeV) of excited states corresponding to the 
$K=1/2$ ground state bands in $^{19}$F and $^{43}$Sc with the 
$-\chi Q\cdot Q$ interaction.
\par
\begin{center}
\begin{tabular}{ccccc}
\multicolumn{2}{c}{$^{19}$F$^{\;a}$} &  & 
\multicolumn{2}{c}{$^{43}$Sc$^{\;b} $} \\ 
$I^{\pi }$ & $E^{*}$ &  & $I^{\pi }$ & $E^{*}$ \\ 
$\left( 1/2\right) ^{+}$ & 0 &  & $\left( 1/2\right) ^{-}$ & 0 \\ 
$\left( 3/2\right) ^{+}$ & 1.588 &  & $\left( 3/2\right) ^{-}$ & 0 \\ 
$\left( 5/2\right) ^{+}$ & 1.588 &  & $\left( 5/2\right) ^{-}$ & 0.679 \\ 
$\left( 7/2\right) ^{+}$ & 5.295 &  & $\left( 7/2\right) ^{-}$ & 0.679 \\ 
$\left( 9/2\right) ^{+}$ & 5.295 &  & $\left( 9/2\right) ^{-}$ & 1.900 \\ 
$\left( 11/2\right) ^{+}$ & 11.118 &  & $\left( 11/2\right) ^{-}$ & 1.900 \\ 
$\left( 13/2\right) ^{+}$ & 11.118 &  & $\left( 13/2\right) ^{-}$ & 3.664 \\ 
&  &  & $\left( 15/2\right) ^{-}$ & 3.664 \\ 
&  &  & $\left( 17/2\right) ^{-}$ & 5.971 \\ 
&  &  & $\left( 19/2\right) ^{-}$ & 5.971
\end{tabular}
\end{center}
\par
$^{a}$ For $^{19}$F we use $\chi =0.1841\;\left( \bar{\chi}=0.0882\right) $
\par
$^{b}$ For $^{43}$Sc we use $\chi =0.0294\;\left( \bar{\chi}=0.0218\right) $
\end{table}

\newpage

\begin{table}[tbp]
{\bf Table II. }$T=0$ and $T=1$ energy levels (MeV) of $^{22}$Na calculated
with the $-\chi Q\cdot Q$ interaction $^{a}$. Only the first six levels for
each $I^{\pi }$ are shown.
\par
\begin{center}
\begin{tabular}{ccccccc}
$I^{\pi }$ & $T=0$ states & $T=1$ states &  & $I^{\pi }$ & $T=0$ states & 
$T=1$ states \\ 
&  &  &  &  &  &  \\ 
$0^{+}$ & 8.999 & \underline{0.000} &  & $6^{+}$ & 7.941 & 10.059 \\ 
& 12.176 & 2.647 &  &  & \underline{11.117} & \underline{11.117} \\ 
& 12.176 & 8.999 &  &  & 11.117 & 11.118 \\ 
& 13.235 & 9.000 &  &  & 14.824 & 16.412 \\ 
& 16.410 & 12.176 &  &  & 16.411 & 16.412 \\ 
& 16.411 & 12.176 &  &  & 16.411 & 16.412 \\ 
&  &  &  &  &  &  \\ 
$1^{+}$ & \underline{0.000} & 2.647 &  & $7^{+}$ & 11.117 & 14.823 \\ 
& 1.588 & 8.999 &  &  & 11.117 & 16.940 \\ 
& 1.588 & 8.999 &  &  & \underline{14.823} & 19.587 \\ 
& 2.647 & 10.059 &  &  & 16.941 & 19.587 \\ 
& 9.000 & 10.059 &  &  & 19.058 & 19.587 \\ 
& 9.000 & 10.059 &  &  & 19.059 & 19.587 \\ 
&  &  &  &  &  &  \\ 
$2^{+}$ & \underline{1.588} & \underline{1.588} &  & $8^{+}$ & 14.823 & 
16.941 \\ 
& 1.588 & 1.588 &  &  & \underline{19.058} & \underline{19.058} \\ 
& 3.176 & 2.647 &  &  & 19.059 & 19.059 \\ 
& 8.999 & 5.294 &  &  & 22.763 & 22.767 \\ 
& 10.059 & 9.000 &  &  & 23.292 & 23.293 \\ 
& 10.059 & 9.000 &  &  & 23.293 & 23.293 \\ 
&  &  &  &  &  &  \\ 
$3^{+}$ & 1.588 & 3.176 &  & $9^{+}$ & 19.058 & 23.822 \\ 
& 1.588 & 5.293 &  &  & 19.058 & 25.939 \\ 
& \underline{3.177} & 10.058 &  &  & \underline{23.822} & 26.470 \\ 
& 5.294 & 10.058 &  &  & 25.940 & 27.527 \\ 
& 5.294 & 11.646 &  &  & 26.469 & 27.527 \\ 
& 5.294 & 11.646 &  &  & 27.528 & 29.644 \\ 
&  &  &  &  &  &  \\ 
$4^{+}$ & 3.176 & \underline{5.293} &  & $10^{+}$ & 23.823 & 25.942 \\ 
& \underline{5.293} & 5.293 &  &  & \underline{29.117} & \underline{29.117}
\\ 
& 5.293 & 5.293 &  &  & 30.705 & 30.706 \\ 
& 7.941 & 10.059 &  &  & 32.293 & 32.294 \\ 
& 11.647 & 11.647 &  &  & 33.881 & 32.294 \\ 
& 11.647 & 11.647 &  &  &  & 33.882 \\ 
&  &  &  &  &  &  \\ 
$5^{+}$ & 5.294 & 7.941 &  &  &  &  \\ 
& 5.294 & 10.059 &  &  &  &  \\ 
& \underline{7.941} & 13.763 &  &  &  &  \\ 
& 10.059 & 13.763 &  &  &  &  \\ 
& 11.118 & 13.763 &  &  &  &  \\ 
& 11.118 & 13.763 &  &  &  & 
\end{tabular}
\end{center}
\par
$^{a}$ In this table and in the following tables, the same value of $\chi $
(and of $\bar{\chi}$) was used for $^{22}$Na as for $^{19}$F.
\end{table}

\newpage

\begin{table}[tbp]
{\bf Table III. }The energies $E^{*}$ (MeV) and percentages of $2\hbar
\omega $ and $4\hbar \omega $ occupancies of the $J=0^{+}\;T=0$ states in 
$^{8}$Be in three different model spaces. The last column gives the \% energy
deviation between the $\left( 0+2\right) ~\hbar \omega $ and $\left(
0+2+4\right) ~\hbar \omega $ calculations.
\par
\begin{center}
\begin{tabular}{cccccccccc}
$0~\hbar \omega $ &  & \multicolumn{2}{c}{$\left( 0+2\right) ~\hbar \omega $}
&  & \multicolumn{3}{c}{$\left( 0+2+4\right) ~\hbar \omega $} &  & \%
deviation \\ \hline
$\chi =0.5216$ &  & \multicolumn{2}{c}{$\chi =0.4119$} &  & 
\multicolumn{3}{c}{$\chi =0.3369$} &  &  \\ 
$E^{*}$ &  & $E^{*}$ & $\%2\hbar \omega $ &  & $E^{*}$ & $\%2\hbar \omega $
& $\%4\hbar \omega $ &  &  \\ 
&  &  &  &  &  &  &  &  &  \\ 
0.00 &  & 0.000 & 21.77 &  & 0.000 & 24.60 & 10.74 &  &  \\ 
9.12 &  & 11.458 & 10.28 &  & 11.370 & 10.90 & 5.69 &  & 0.77 \\ 
12.16 &  & 16.231 & 1.91 &  & 15.879 & 2.07 & 3.55 &  & 2.22 \\ 
15.20 &  & 18.352 & 2.96 &  & 17.860 & 2.53 & 3.16 &  & 2.75 \\ 
17.23 &  & 20.147 & 2.62 &  & 19.385 & 2.10 & 2.98 &  & 3.93 \\ 
&  & 30.951 & 88.89 &  & 26.225 & 50.93 & 20.56 &  & 18.02 \\ 
&  & 32.214 & 100.00 &  & 29.701 & 77.30 & 19.41 &  & 8.46 \\ 
&  & 32.396 & 92.36 &  & 32.085 & 86.13 & 13.87 &  & 0.97 \\ 
&  & 34.409 & 100.00 &  & 34.200 & 86.82 & 13.18 &  & 0.61 \\ 
&  & 38.702 & 91.66 &  & 35.928 & 70.75 & 15.45 &  & 7.72
\end{tabular}
\end{center}
\end{table}

\newpage

\begin{table}[tbp]
{\bf Table IV. }Same as Table III but for the $J=1^{+}$ states.
\par
\begin{center}
\begin{tabular}{cccccccccc}
$0~\hbar \omega $ &  & \multicolumn{2}{c}{$\left( 0+2\right) ~\hbar \omega $}
&  & \multicolumn{3}{c}{$\left( 0+2+4\right) ~\hbar \omega $} &  & \%
deviation \\ \hline
$\chi =0.5216$ &  & \multicolumn{2}{c}{$\chi =0.4119$} &  & 
\multicolumn{3}{c}{$\chi =0.3369$} &  &  \\ 
$E^{*}$ &  & $E^{*}$ & $\%2\hbar \omega $ &  & $E^{*}$ & $\%2\hbar \omega $
& $\%4\hbar \omega $ &  &  \\ 
&  &  &  &  &  &  &  &  &  \\ 
9.12 &  & 11.458 & 10.28 &  & 11.371 & 10.90 & 5.69 &  & 0.77 \\ 
11.15 &  & 13.734 & 8.85 &  & 13.587 & 8.92 & 4.87 &  & 1.08 \\ 
15.20 &  & 18.353 & 2.96 &  & 17.861 & 2.53 & 3.16 &  & 2.75 \\ 
17.23 &  & 20.148 & 2.62 &  & 19.385 & 2.10 & 2.98 &  & 3.94 \\ 
&  & 30.722 & 100.00 &  & 27.896 & 78.69 & 21.31 &  & 10.13 \\ 
&  & 32.992 & 100.00 &  & 30.215 & 79.30 & 20.70 &  & 9.19 \\ 
&  & 33.294 & 100.00 &  & 31.756 & 82.55 & 17.45 &  & 4.84 \\ 
&  & 34.409 & 100.00 &  & 33.731 & 86.28 & 13.72 &  & 2.01 \\ 
&  & 34.503 & 100.00 &  & 33.869 & 83.31 & 16.59 &  & 1.87 \\ 
&  & 35.357 & 99.84 &  & 34.200 & 86.81 & 13.19 &  & 3.38 \\ 
&  & 37.666 & 92.98 &  & 35.707 & 74.99 & 14.33 &  & 5.49 \\ 
&  & 38.702 & 91.65 &  & 35.927 & 70.75 & 15.44 &  & 7.72 \\ 
&  & 40.174 & 100.00 &  & 38.724 & 85.12 & 14.80 &  & 3.74 \\ 
&  & 40.485 & 99.97 &  & 39.313 & 86.37 & 13.63 &  & 2.98 \\ 
&  & 40.706 & 100.00 &  & 39.331 & 86.38 & 13.63 &  & 3.50
\end{tabular}
\end{center}
\end{table}

\newpage

\begin{table}[tbp]
{\bf Table V. }Same as Table III but for the $J=2^{+}$ states.
\par
\begin{center}
\begin{tabular}{cccccccccc}
$0~\hbar \omega $ &  & \multicolumn{2}{c}{$\left( 0+2\right) ~\hbar \omega $}
&  & \multicolumn{3}{c}{$\left( 0+2+4\right) ~\hbar \omega $} &  & \%
deviation \\ \hline
$\chi =0.5216$ &  & \multicolumn{2}{c}{$\chi =0.4119$} &  & 
\multicolumn{3}{c}{$\chi =0.3369$} &  &  \\ 
$E^{*}$ &  & $E^{*}$ & $\%2\hbar \omega $ &  & $E^{*}$ & $\%2\hbar \omega $
& $\%4\hbar \omega $ &  &  \\ 
&  &  &  &  &  &  &  &  &  \\ 
3.04 &  & 3.042 & 21.42 &  & 3.038 & 23.81 & 9.94 &  & 0.13 \\ 
9.12 &  & 11.458 & 10.28 &  & 11.370 & 10.90 & 5.69 &  & 0.77 \\ 
11.15 &  & 13.734 & 8.85 &  & 13.586 & 8.92 & 4.87 &  & 1.09 \\ 
12.15 &  & 16.197 & 8.99 &  & 15.880 & 2.07 & 3.55 &  & 2.00 \\ 
14.18 &  & 16.231 & 1.91 &  & 15.949 & 8.31 & 4.21 &  & 1.77 \\ 
15.20 &  & 18.353 & 2.96 &  & 17.859 & 2.53 & 3.16 &  & 2.77 \\ 
17.23 &  & 20.148 & 2.62 &  & 19.385 & 2.10 & 2.98 &  & 3.94 \\ 
&  & 31.711 & 82.81 &  & 27.146 & 51.37 & 20.15 &  & 16.82 \\ 
&  & 32.214 & 100.00 &  & 30.215 & 79.30 & 20.70 &  & 6.62 \\ 
&  & 32.992 & 100.00 &  & 31.714 & 80.12 & 18.91 &  & 4.03 \\ 
&  & 33.962 & 99.68 &  & 32.085 & 86.13 & 13.87 &  & 5.85 \\ 
&  & 34.410 & 100.00 &  & 33.869 & 83.31 & 16.60 &  & 1.60 \\ 
&  & 35.357 & 99.84 &  & 34.201 & 86.81 & 13.19 &  & 3.38 \\ 
&  & 37.668 & 92.98 &  & 35.707 & 75.00 & 14.33 &  & 5.49
\end{tabular}
\end{center}
\end{table}

\end{document}